\documentclass[superscriptaddress,twocolumn]{revtex4}
\usepackage{mathrsfs}
\usepackage{amssymb}
\usepackage[tbtags]{amsmath}
\usepackage{graphicx}
\usepackage{epsfig,graphicx,times}
\usepackage{graphicx}
\usepackage{subfigure}
\usepackage{color}

\begin{document}
\title{Quantum computation with moving quantum dots generated by surface acoustic waves}

\author{X. Shi}
\affiliation{Quantum Optoelectronics Laboratory, School of Physics
and Technology, Southwest Jiaotong University, Chengdu 610031,
China}

\author{M. Zhang}
\affiliation{Quantum Optoelectronics Laboratory, School of Physics
and Technology, Southwest Jiaotong University, Chengdu 610031,
China}

\author{L. F. Wei\footnote{weilianfu@gmail.com}}
\affiliation{Quantum Optoelectronics Laboratory, School of Physics
and Technology, Southwest Jiaotong University, Chengdu 610031,
China} \affiliation{State Key Laboratory of Optoelectronic Materials
and Technologies, School of Physics Science and Engineering, Sun
Yet-sen University, Guangzhou 510275, China}

\date{\today}
\begin{abstract}
Motivated by the recent experimental observations [M. Kataoka et
al., Phys. Rev. Lett. {\bf102}, 156801 (2009)], we propose here an
theoretical approach to implement quantum computation with bound
states of electrons in moving quantum dots generated by the driving
of surface acoustic waves.
Differing from static quantum dots defined by a series of static
electrodes above the two-dimensional electron gas (2DEG), here a
single electron is captured from a 2DEG-reservoir by a surface
acoustic wave (SAW) and then trapped in a moving quantum dot (MQD)
transporting across a quasi-one dimensional channel (Q1DC), wherein
all the electrons have been excluded out by the actions of the
surface gates.
The flying qubit introduced here is encoded by the two lowest
adiabatic levels of the electron in the MQD, and the Rabi
oscillation between these two levels could be implemented by
applying finely-selected microwave pulses to the surface gates. By
using the Coulomb interaction between the electrons in different
moving quantum dots, we show that a desirable two-qubit operation,
i.e., i-SWAP gate, could be realized.
Readouts of the present flying qubits are also feasible with the
current single-electron detected technique.


\end{abstract}
\maketitle
\section{Introduction}
%\begin{flushleft}
%\textbf{1. }
%\end{flushleft}
In the past decades a considerable attention is paid to quantum
computation implemented usually by an array of weakly-coupled
quantum systems~\cite{Lloyd}. Basically, a quantum computing process
involves a series of time-evolution of the coupled two-level quantum
systems (qubits). In the classical computer, the information unit is
represented by a bit, which is always understood as either $0$ or
$1$. The information unit in quantum computation is very different.
For example, the qubit can be at logic ``0" or logic ``1" and also
the superposition of both. Owing to this property, quantum computer
provides an automatically-parallel computing and thus possesses much
more powerful features than that realized by the classical computer.
This basic advantage has been definitely demonstrated with Shor
algorithm~\cite{Shor} for significantly speeding up the large number
factoring.

A central challenge in the current quantum information science is,
how to build such a quantum computer? Until now, there has been many
proposals for experimental quantum computation, such as atomic
qubits coupled via a cavity field~\cite{Parkins,Pellizzari}, cold
ions confined in a linear trap~\cite{CiracN}, nuclear magnetic
resonance~\cite{Gershenfeld}, photons~\cite{Bouwmeester,Kim},
quantum dots~\cite{Loss,Brown}, and Josephson superconducting
system~\cite{SC}, etc.. Note that all these candidates are based on
the static qubits, and the controllable interbit interactions are
difficult to achieve. Alternatively, in this paper we focus on the
flying qubits generated by the electrons in moving quantum dots
(MQDs). In fact, quantized transport of electrons along a quasi-one
dimensional channel (Q1DC) by surface acoustic waves have been
observed~\cite{Shilton,Talyanskii}.
The original attempt in these experiments is to build the desirable
current standards, but now has also leaded to the study for quantum
computation. The qubit in such a systems is
"flying"~\cite{Barnes,Furuta}, since the electron in the MQD is
drawn along the Q1DC by a surface acoustic wave (SAW). In principle,
quantum computing with these flying qubits realized by using SAWs
possess two manifest advantages~\cite{Roberta,Bordone}; i) one can
make ensemble measurements over billions of identical MQDs and thus
be robust against various random errors, and ii) it should allow a
longer quantum operation by preventing the spreading of the wave
function and reducing undesired reflection effects.

The approach using the above SAW-based flying qubits to implement
quantum computation was first proposed by Barnes et
al~\cite{Barnes}, who used two spin-states of the transported
electron to encode a flying qubit. Although the feasibility of this
proposal was then analyzed in detail~\cite{Furuta}, the experimental
demonstration of this proposal has not been achieved yet.
One of the possible obstacles is that the required local magnetic
fields are not easy to realize for manipulating the spin-states of
the electrons in the MQDs. In order to overcome such a difficulty,
the flying qubits in our quantum computing proposal are directly
encoded by the two bound-states (rather than the above spin-states)
of the electrons in the MQDs. Our idea is motivated by the recent
experimental work, wherein the coherent single-electron dynamics on
these bound states was successfully observed~\cite{KataokaL}.

The paper is organized as follows. In Sec. II we briefly describe
the SAW-based MQDs and numerically calculate the electronic levels.
By applying an additional driving electric field to the gates above
the channel, we show in Sec. III that the Rabi oscillations between
the qubit's levels could be implemented. In Sec. IV, we describe an
approach to implement a two-qubit operation between the flying
qubits across different channels. Finally, we summarize our main
results and give some discussions on feasibility of our proposal,
including how to read out the proposed flying qubit by using the
existing experimental-technique.

\section{SAW-based moving quantum dots}
We consider a system showed in Fig.~1
~\cite{Cunningham,KataokaL,Kataoka,Robinson}, wherein
quantized-acoustoelectric-current driven by SAW was observed.
A two-dimensional electron gas (2DEG) is formed in a GaAs/AlGaAs
heterostructure below the metallic surface split-gate. At $1.5~K$
the electronic density and the mobility in this 2DEG are measured
as~\cite{Kataoka} $1.8\times 10^{15}~ m^{-2}$ and $160
~m^{2}V^{-1}s^{-1}$, respectively.
The surface gates are utilized to define a Q1DC without any
electron. Two SAW interdigital transducers placed on each side of
the device are used to generate a SAW (with a resonant frequency
around $3~ GHz$) propagating along the Q1DC. The surface gate
geometry is chosen to produce an electrostatically defined channel
with the length approximately of the SAW wavelength
($\lambda=1~\mu\text{m}$), so that a single electron can be
periodically transported through the channel. The moving potential
containing few electrons related to the SAW's propagation can be
considered as a MQD.
Of course, when the quantum dot carrying few electrons moves through
the channel, a quantized current is generated. This current can be
measured by connecting an ammeter to two Ohmic contacts on the 2DEG
mesa.

\begin{figure}[htbp]
\includegraphics[width=8cm,height=5cm]{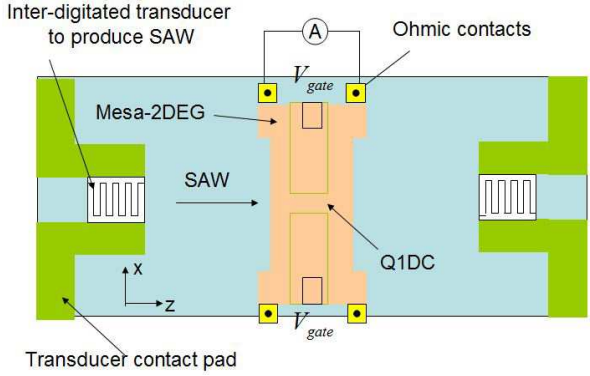}
\caption{A schematic diagram of the MQD
device~\cite{Cunningham,KataokaL,Kataoka,Robinson}.}
\end{figure}

For simplicity, we assume that only one electron is captured into
the MQD and then propagates along the narrow depleted Q1DC.
The potential of the electron in a MQD could be effectively
simplified as
\begin{equation}
V_{\rm eff}(z, t)=V_{\rm SAW}(z, t)+V_{\rm gate}(z),
\end{equation}
where $V_{\rm SAW}(z, t)$ and $V_{\rm gate}(z)$ are the
piezoelectric potential accompanying the SAW and the electrostatic
potential defined by the surface split-gate, respectively.
First, the thickness and width of the quantum dot (i.e., its sizes
along the $y$- and $x$-direction) are all neglected, such that the
electrostatic potential could be simply modeled as a strictly 1D
potential~\cite{GodfreyB,Godfrey}
\begin{align}
V_{\rm gate}(z)=\frac{V_0}{\cosh^{2}(z/a)}.\label{eq:1}
\end{align}
Here, the $z$-axis is chosen along the channel which the SAW
propagates through and the parameter $V_0$ determines the effective
height of the potential barrier.
The split-gate is operated well beyond the pinch off voltage in the
absence of the SAW, so the energy $V_0$ could be greater than the
electron Fermi energy in the 2DEG, and the edge of the depleted Q1DC
is well away from the edge of the surface split-gate. The effective
length of the Q1DC can be taken as $l_{\rm eff}=2a$, and it takes
also approximately as long as the SAW wavelength $\lambda=1\mu$m.
Consequently, we have $a=0.5~\mu$m.
Next, by considering the screening effect of the metal gates on the
SAW-induced electric potential, and neglecting the mechanical
coupling between the semiconductor and the metal surface gate, all
the changes in the components of the stress tensor, and the
separation between split-gates, etc., A\v{\i}zin
$et~al$~\cite{Aizin} showed that the piezoelectric potential $V_{\rm
SAW}$ could be simplified to the form
\begin{align}
V_{\rm SAW}=V_S\cos(kz-wt).\label{eq:2}
\end{align}
Here, $V_S$ is the amplitude of the SAW, and $k$ and $w$ are the
frequency and wave number, respectively.

With the above potential the electronic levels of the electron
trapped in the MQD can be determined by solving the instantaneous
eigenvalue equation
\begin{align}
&\hat{H}_0(t)\rvert E_n(t)\rangle=E_n(t)\rvert \psi_n(t)\rangle,\nonumber\\
&\hat{H}_0(t)=-\frac{\hbar^2}{2m^*}\frac{d^2}{dz^2}+\frac{V_0}{\cosh^{2}(z/a)}
+V_S\cos(kx-wt).\label{eq:4}
\end{align}
Here, $m^*=0.0067m_e$ is the effective mass of the electron in GaAs,
and $V_0=\hbar^2/2m^* l_0^2$, $V_S=\gamma V_0$. The parameter
$l_0=4\times 10^{-2}a$ is the effective width of the Q1DC, and
$\gamma=0.5$ the ratio of the SAW potential amplitude to the height
of the electrostatically-induced potential barrier in the Q1DC. The
SAW velocity is $v=2981$m/s~\cite{GodfreyB}.
By finite differential method we can numerically solve Eq.
\eqref{eq:4} and obtain the electronic levels in the MQD. Although
the shape of the potential or the size of the quantum dot changes
with the motion, the dot is still ``big" enough to hold a few
levels.
Specifically, Fig.~2 shows the effective potential and its
corresponding bound levels for the different times over the SAW
period. Qualitatively, the dot could capture many electrons
initially, but most of them will be escaped from the local well and
returned to the source reservoir. In the present calculation, we
consider the ideal condition wherein only one electron is initially
captured by the MQD and held in where across the channel.
One can see from Fig.~2 that, a few bound levels exist in the local
potential of the quantum dot moving along the channel. The wave
function and the corresponding probabilistic distributions of the
electron residing in these levels are shown in Fig.~3. One can see
that the electron in the third (blue-line) level or the higher ones
could escape from the well. While, the probabilities of the electron
in the lowest two levels, the ground and first excited ones,
tunneling to the source reservoir is negligible. As a consequence,
these two levels can be used to encode the desirable flying qubit,
the unit of the moving quantum information.

\begin{figure}[htbp]
\includegraphics[width=8cm,height=10cm,angle=270]{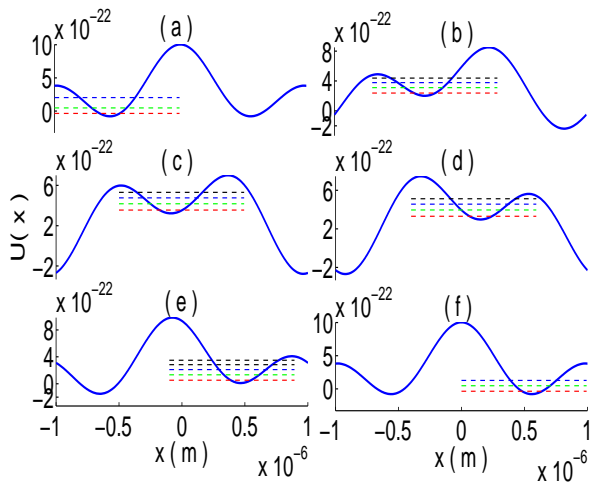}
\caption{The effective potential and its allowed energy levels for
$\gamma=0.5$ at various typical times: (a) $t=0$, (b) $t=0.1$ns, (c)
$t=0.15$ns, (d) $t=0.2$ns, (e) $0.3\times 10^{-9}~s$, and (f)
$t=T=0.34$ns. The blue solid line represent the effective potential
and the colored dashed-lines show the allowed levels: red (ground
state), green (the first exited state), blue (the second excited
state), etc..}
\end{figure}

\begin{figure}[htbp]
\includegraphics[width=9cm,height=9cm,angle=270]{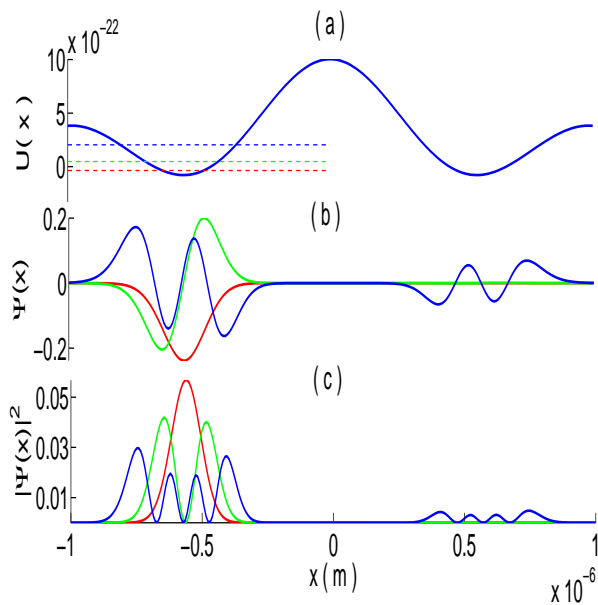}
\caption{Wave functions of the lowest three levels and their
relevant probabilistic distributions at certain time: (a) Potential
and its allowed levels, (b) The eigenfunctions of the allowed levels
and (c) the probabilistic distributions of the electron in the
allowed-levels along the channel. Here, the red, green and blue
lines denote the ground, the first excited and the second excited
state, respectively.}
\end{figure}

We now show that the flying qubit defined above is sufficiently
stationary, although the shape of the potential varies with the
quantum dot moving along the channel. The adiabatic theorem asserts
that, if the rate of the change of Hamiltonian is slow enough, the
system will stay at an instantaneous eigenstate of the
time-Hamiltonian. For the present case the adiabatic condition is
expressed as
\begin{align}
\beta=\left| \frac{\hbar \langle m\lvert \frac{\partial
H_0(t)}{\partial t}\rvert n \rangle }{(E_m(t)-E_n(t))^2}\right| \ll
1, \label{eq:5}
\end{align}
where $E_m(t)-E_n(t)$ is the energy splitting between the state
$\rvert m \rangle$ and $\rvert n \rangle$. Our numerical results
show that, at certain time: $E_0=-3.57\times 10^{-23}~J,
E_1=4.80\times 10^{-23}~J$ and consequently $\beta=0.0289$. This
indicates that the adiabatic condition could be satisfied. Less
value of the $\beta$-parameter is also possible by properly
adjusting the relevant parameters. This means that the levels used
above to encode the flying qubit is adiabatic. Thus, once the flying
qubit is prepared at one of its logic states ($|0\rangle$ and
$|1\rangle$), it always stays at that state until the specific
driving is applied.

\section{Rabi oscillations between the levels of flying qubit}
For realizing quantum computation, we need to first implement
arbitrary rotations of the single-qubit. For the present flying
qubit, this can be achieved by using the usual Rabi oscillations
between the adiabatic states $\rvert 0 \rangle$ and $\rvert 1
\rangle$.
Basically, these states should be kept as the pure ones. This can be
realized by cooling the system to a sufficiently low temperature
$T_{\rm temp}$, such that the condition $k_B T_{\rm temp}\ll
\hbar\omega$ is satisfied. Here, $\omega=\omega_1-\omega_0$ is the
electronic transition frequency of the flying qubit, and $k_B$ is
Boltzmann constant. Experimentally~\cite{KataokaL}, the system can
be worked approximately at the temperature $T_{\rm temp}=0.27~K$,
yielding $k_BT_{\rm temp}=3.726\times 10^{-24}~J\ll \hbar\omega\sim
8.3667\times 10^{-23}~J$. Thus, the transitions between the qubit's
levels due to the thermal excitations can be safely neglected.

We now apply a resonant electric driving to the surface gates for
implementing the desirable Rabi oscillations. Under such a driving
the previous 1D-potential $V_{\rm gate}$, i.e., Eq. (2), is now
changed as $V_{\rm gate}\rightarrow V'_{\rm gate}=V_{\rm
gate}+V_e\cos(\omega t)/\cosh^2(z/a)$. Consequently, the dynamics of
the driven flying qubit is determined by the following
time-dependent Schr\"{o}dinger equation
\begin{equation}
i\hbar \frac{\partial\rvert \psi(t) \rangle}{\partial
t}=(\hat{H}_0+\hat{H}')\rvert \psi(t) \rangle. \label{eq:7}
\end{equation}
Here,
\begin{equation}
\hat{H}'=\frac{V_e\cos(\omega t)}{\cosh^{2}(z/a)},\label{eq:8}
\end{equation}
describes the driving induced by the applied oscillating
electric-field, which is perpendicular to the Q1DC and linearly
polarized along the x-axis. Above, $V_e$ is a parameter depending on
the power of the applied electric-filed. In our calculation, we
choose it as $V_e=0.1V_S$ for simplicity.
Generally, the wave function of the driven flying qubit can be
written as
\begin{equation}
\rvert \psi(t) \rangle =C_0 \rvert 0 \rangle +C_1 \rvert 1
\rangle,\label{eq:6}
\end{equation}
with $C_0$ and $C_1$ being the probability-amplitudes of finding the
electron in the states $\rvert 0 \rangle$ and $\rvert 1 \rangle$,
respectively.

From Eqs.~(6-8), the equations of motion for the amplitudes $C_0$
and $C_1$ can be derived as
\begin{equation}
\frac{\partial C_0}{\partial t}=-i\omega_0C_0-iC_0D_{00}\cos{(\omega
t)}-iC_1D_{01}\cos(\omega t),\label{eq:9}
\end{equation}
and
\begin{equation}
\frac{\partial C_1}{\partial t}=-i\omega_1C_1-iC_1D_{11}\cos{(\omega
t)}-iC_0D_{10}\cos(\omega t),\label{eq:10}
\end{equation}
with $D_{ij}=V_e/[\hbar\langle i|\cosh^2(z/a)|j\rangle],\,i, j=0,1$.
The above equations can be exactly solved by numerical method. Then,
the time-dependent probabilities of the electron being in the states
$\lvert 0 \rangle$ and $\lvert 1 \rangle $ can be obtained as
$\lvert C_0(t) \rvert ^2$ and $\lvert C_1(t) \rvert ^2$,
respectively. Certainly, the relation
\begin{align}
\lvert C_0(t) \rvert ^2+\lvert C_1(t) \rvert ^2=1,\label{eq:11}
\end{align}
is always satisfied. With the initial condition
$|\psi(0)\rangle=|0\rangle$ we plot the time-dependent $|C_1(t)|^2$
in Fig.~4.
It is seen really that the population in one of the logic state of
the flying reveals a obviously oscillating behavior with a period:
$\tau\sim 0.32$~ns for the parameters selected above. This
time-interval is sufficiently-long for the MQD across the Q1DC
demonstrated in the experiment. The time interval for a quantum dot
across the channel is estimated as $\sim 0.34~ns$. Thus, Rabi
oscillations can be really utilized to realize the desirable
single-qubit operations.

\begin{figure}[htbp]
\includegraphics[width=6cm,height=8cm,angle=270]{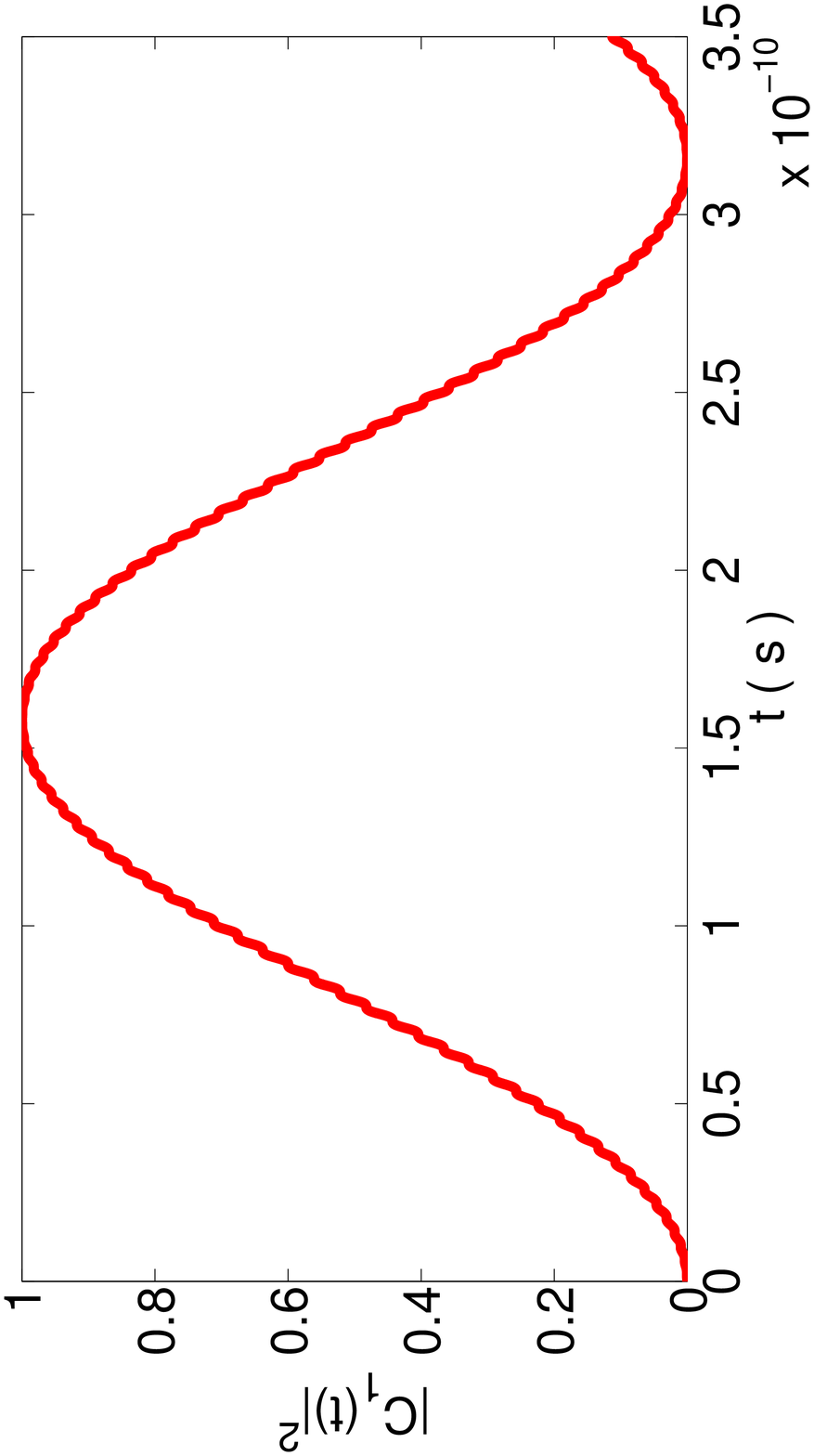}
\caption{Rabi oscillation of the population in flying qubit's level
$|1\rangle$. The oscillation period shown here is about $0.32$~ns.}
\end{figure}

\section{Coupling the separated moving quantum dots
for two-flying-qubit operations}
\begin{figure}[htbp]
\includegraphics[width=8cm,height=4cm]{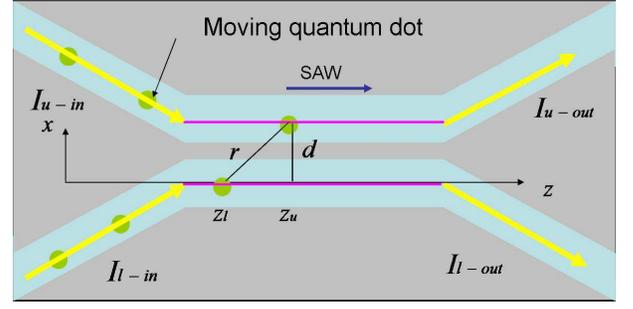}
\caption{The schematic diagram to implement controllable couplings
between two flying-qubits. Two MQDs passage along the the upper- and
lower Q1DCs, respectively and the coupling between them is realized
by the Coulomb interaction of the inside electrons.}
\end{figure}
We now discuss how to implement an universal gate, i.e., the
two-qubit operation, with the MQDs. A simple way to achieve such a
task is by utilizing the Coulomb interaction between the electrons
in the nearest-neighbour interaction MQDs. To do this, let us
consider the situation schematically shown in Fig.~5, wherein two
MQDs driven by two SAWs pass across two Q1DCs, the upper- and lower
ones. Suppose that the tunneling between them is negligible and only
the Coulomb interaction between them is important. First, the
Coulomb force between the electrons in these two MQDs can be
expressed as
\begin{align}
F_{\rm int}(z_u, z_l)
=\frac{e^2}{4\pi\varepsilon_0}\frac{(z_l-z_u)}{[d^2+(z_l-z_u)^2]^{3/2}},\label{eq:12}
\end{align}
with $z_u$ and $z_l$ being their coordinates along the channels (the
indices $u$ and $l$ refer to the upper and lower channels,
respectively) and $d$ the distance between the two Q1DCs.
Since the motions of the electrons are always along the Q1DCs, the
vertical force of the Coulomb interaction can be ignored and thus
only the horizontal force along the z-axis is taken account into.
Second, the potential related to above force can be written as
\begin{equation}
\begin{array}{l}
V_{\rm
int}(z)=\frac{1}{4\pi\varepsilon_0}\int_{0}^{z}\frac{e^2zdz}{(d^2+z^2)^{3/2}}
\\
\\
\,\,\,\,\,\,\,\,\,\,\,\,\,\,\,
\,\,\,\,=-\frac{1}{4\pi\varepsilon_0}\frac{e^2}{d}\left\{\frac{1}{[1+(z/d)^2]^{1/2}}-1\right\},\label{eq:13}
\end{array}
\end{equation}
where $z=z_l-z_u$.
By using the usual Taylor expansion and ignoring the high-order
terms under the condition $d\gg z$, the above Coulomb potential
reduces to
\begin{align}
V_{\rm
int}(z)=\frac{e^2}{8\pi\varepsilon_0d^3}z^2=\frac{e^2}{8\pi\varepsilon_0d^3}\left(z_u^2+z_l^2-2z_u
z_l\right).\label{eq:14}
\end{align}
Thirdly, the Hamiltonian describing the dynamics of the two-coupling
MQDs reads
\begin{align}
\hat{H}_h&=\hat{H}_t+V_{\rm int}(z),\label{eq:15}
\end{align}
with
\begin{equation}
\begin{array}{l}
\hat{H}_t=\sum_{j=u,l}\left[-\frac{\hbar^2}{2m_j^*}\frac{d^2}{dz_j^2}
+\frac{V_0^j}{\cosh^{2}(z_j/a)}+V_S^j\cos(kz_j-\omega t)\right]
\\
\\
\,\,\,\,\,\,\,\,\,
=\sum_{j=u,l}\frac{\hbar\omega_j}{2}\hat{\sigma}_j^z,\label{eq:16}
\end{array}
\end{equation}
and $\hat{\sigma}_j^z= \rvert 1_j\rangle \langle 1_j\lvert-\rvert
0_j\rangle \langle 0_j\lvert$, $\omega_j=(E_1^j-E_0^j)/\hbar$.

In the qubit representation, the position operators $\hat{z}_j$,
$\hat{z}_j^2$ and $\hat{z}_j\hat{z}_k$ (where $j,k=u,l$ and $j\neq
k$) can be expressed as
\begin{equation}
\begin{array}{l}
\hat{z}_j=\frac{1}{2}(z_j^{11}-z_j^{00})\hat{\sigma}_j^z+z_j^{01}
\hat{\sigma}_j^x,\\\\
\hat{z}_j^2=\frac{1}{2}(z_j^{11}+z_j^{00})(z_j^{11}-z_j^{00})
\hat{\sigma}_j^z\\\\
\,\,\,\,\,\,\,\,\,\,\,\,
+(z_j^{00}+z_j^{11})z_j^{01}\hat{\sigma}_j^x,\label{eq:19}
\end{array}
\end{equation}
and
\begin{equation}
\begin{array}{l}
\hat{z}_j\hat{z}_k=\frac{1}{4}(z_k^{11}+z_k^{00})(z_j^{11}-z_j^{00})
\hat{\sigma_j^z}+z_j^{01}z_k^{01}\hat{\sigma}_j^x\hat{\sigma}_k^x
\\
\\
\,\,\,\,\,\,\,\,\,\,\,\,\,\,\,\,\,\,
+\frac{1}{4}(z_j^{11}+z_j^{00})(z_k^{11}-z_k^{00})
\hat{\sigma_k^z}
\\
\\
\,\,\,\,\,\,\,\,\,\,\,\,\,\,\,\,\,\,
+\frac{1}{2}(z_k^{00}+z_k^{11})z_j^{01}
\hat{\sigma_j^x}
+\frac{1}{2}(z_j^{00}+z_j^{11})z_k^{01}\hat{\sigma_k^x}
\\
\\
\,\,\,\,\,\,\,\,\,\,\,\,\,\,\,\,\,\,
+\frac{1}{4}(z_j^{11}-z_j^{00})(z_k^{11}-z_k^{00})
\hat{\sigma_j^z}\hat{\sigma_k^z}
\\
\\
\,\,\,\,\,\,\,\,\,\,\,\,\,\,\,\,\,\,
+\frac{1}{2}(z_j^{11}-z_j^{00})z_k^{01}\hat{\sigma}_j^z
\hat{\sigma}_k^x+\frac{1}{2}(z_k^{11}-z_k^{00})z_j^{01}\hat{\sigma}_j^z\hat{\sigma}_k^x
,
\end{array}
\end{equation}
respectively. Above, $\hat{\sigma}_j^x=\hat{\sigma}_j^+
+\hat{\sigma}_j^-$ with $\hat{\sigma}_j^+=\rvert 1_j \rangle \langle
0_j\lvert$ and $\hat{\sigma}_j^-=\rvert 0_j \rangle \langle
1_j\lvert$; $z_j^{11}$, $z_j^{00}$ and $z_j^{01}$ are the matrix
elements $\langle 1_j \lvert z_j \rvert 1_j \rangle$, $\langle 0_j
\lvert z_j \rvert 0_j \rangle$, and $\langle 1_j \lvert z_j \rvert
0_j \rangle$, respectively. As a consequence, the above Coulomb
potential $V_{\rm int}(z)$ can be rewritten as
\begin{equation}
\begin{array}{l}
\hat{V}_{\rm int}=C_u^z\hat{\sigma}_u^z+C_l^z\hat{\sigma}_l^z+C_u^x
\hat{\sigma}_u^x+C_l^x\hat{\sigma}_l^x
+C_{ul}^{zz}\hat{\sigma}_u^z\hat{\sigma}_l^z
\\
\\
\,\,\,\,\,\,\,\,\,\,\,\,\,\,\,\,
+C_{ul}^{xx}\hat{\sigma}_u^x\hat{\sigma}_l^x
+C_{ul}^{zx}\hat{\sigma}_u^z\hat{\sigma}_l^x+C_{ul}^{xz}
\hat{\sigma}_u^x\hat{\sigma}_l^z,\label{eq:21}
\end{array}
\end{equation}
with
\begin{eqnarray}
\left\{
\begin{array}{l}
C_u^z=\frac{e^2}{16\pi\varepsilon_0d^3}(z_u^{00}+z_u^{11}-z_l^{00}-z_l^{11})(z_u^{11}-z_u^{00}),
\\\\
C_l^z=\frac{e^2}{16\pi\varepsilon_0d^3}(z_l^{00}+z_l^{11}-z_u^{00}-z_u^{11})(z_l^{11}-z_l^{00}),
\\\\
C_u^x=\frac{e^2}{8\pi\varepsilon_0d^3}(z_u^{00}+z_u^{11}-z_l^{00}-z_l^{11})z_u^{01},
\\\\
C_l^x=\frac{e^2}{8\pi\varepsilon_0d^3}(z_l^{00}+z_l^{11}-z_u^{00}-z_u^{11})z_l^{01},
\end{array}
\right.
\end{eqnarray}
and

\begin{eqnarray}
\left\{
\begin{array}{l}
C_{ul}^{zz}=\frac{-e^2}{16\pi\varepsilon_0d^3}(z_u^{11}-z_u^{00})(z_l^{11}-z_l^{00}),
\\\\
C_{ul}^{xx}=\frac{-e^2}{4\pi\varepsilon_0d^3}z_u^{01}z_l^{01},
\\\\
C_{ul}^{zx}=\frac{-e^2}{8\pi\varepsilon_0d^3}(z_u^{11}-z_u^{00})z_l^{01},
\\\\
C_{ul}^{xz}=\frac{-e^2}{8\pi\varepsilon_0d^3}(z_l^{11}-z_l^{00})z_u^{01}.
\end{array}
\right.
\end{eqnarray}
In the interaction picture defined by the unitary
$\hat{U}_0(t)=\exp{[(-i/\hbar)t\sum_{j=u,l}\lambda_j\hat{\sigma}_j^z]}$
with $\lambda_j=\hbar\omega_j/2+C_j^z$, the Hamiltonian of the
system reduces to
\begin{equation}
\begin{array}{l}
\hat{H}_{I}=C_{ul}^{zz}\hat{\sigma}_u^z\hat{\sigma}_l^z+\sum_{j=u,l}C_j^x
\left(e^{2it\lambda_j/\hbar}\hat{\sigma}_j^+
+ e^{-2it\lambda_j/\hbar}\hat{\sigma}_j^-\right)
\\\\
\,\,\,\,\,\,\,\,\,\,\,\,\,\,\,\,
+C_{ul}^{xx}[e^{2it(\lambda_u+\lambda_l)/\hbar}\hat{\sigma}_u^+\hat{\sigma}_l^+
+e^{2it(\lambda_u-\lambda_l)/\hbar}\hat{\sigma}_u^+\hat{\sigma}_l^-
\\\\
\,\,\,\,\,\,\,\,\,\,\,\,\,\,\,\,
+e^{-2it(\lambda_u-\lambda_l)/\hbar}\hat{\sigma}_u^-\hat{\sigma}_l^+
+e^{-2it(\lambda_u+\lambda_l)/\hbar}\hat{\sigma}_u^-\hat{\sigma}_l^-]
\\\\
\,\,\,\,\,\,\,\,\,\,\,\,\,\,\,\,
+C_{ul}^{zx}\left(e^{2it\lambda_l/\hbar}\hat{\sigma}_u^z\hat{\sigma}_l^+
+e^{-2it\lambda_l/\hbar}\hat{\sigma}_u^z\hat{\sigma}_l^-\right)
\\\\
\,\,\,\,\,\,\,\,\,\,\,\,\,\,\,\,
+C_{ul}^{xz}\left(e^{2it\lambda_u/\hbar}\hat{\sigma}_u^+\hat{\sigma}_l^z
+e^{-2it\lambda_u/\hbar}\hat{\sigma}_u^-\hat{\sigma}_l^z\right).\label{eq:24}
\end{array}
\end{equation}
Consequently, under the usual rotating-wave approximation, we have
\begin{align}
\overline{H}_{I}=C_{ul}^{xx}(\hat{\sigma}_u^+\hat{\sigma}_l^-
            +\hat{\sigma}_u^-\hat{\sigma}_l^+).\label{eq:25}
\end{align}
During this derivation, the significantly-small quantities
$C_{ul}^{zz}\ll C_{ul}^{xx}$ has been omitted, and we have also
assumed that $\lambda_u=\lambda_l$. Typically, for the experimental
parameters: $z=2.981\times 10^{-2}\mu$m, we have
$z_u^{00}=-5.6186\times 10^{-1}\mu\text{m},\,z_u^{11}=-5.6975\times
10^{-1}\mu\text{m},\,z_u^{01}=z_u^{10}=-5.6431\times
10^{-2}\mu\text{m}$; $z_l^{00}=-5.3594\times
10^{-1}\mu\text{m},\,z_l^{11}=-5.4418\times
10^{-1}\mu\text{m},\,z_l^{01}=z_l^{10}=-5.6607\times
10^{-2}\mu\text{m}$, and thus $C_{ul}^{zz}/C_{ul}^{xx}=1.3\times
10^{-3}\ll 1$.

Finally, the above Hamiltonian yield the following two qubit
evolution (in the representation with the basis$\{\rvert 11 \rangle,
\rvert 10 \rangle,\rvert 01 \rangle,\rvert 00 \rangle\}$
\begin{align}
\hat{U}=e^{-i\hat{H}_It/\hbar}=\left(\begin{array}{cccc}
1         &0      &0           &0\\
0       &\cos{\xi}         &-i\sin{\xi}           &0\\
0        &-i\sin{\xi}       &\cos{\xi}          &0\\
0       &0         &0           &1
\end{array}\right),\,\,\xi=tC_{ul}^{xx}/\hbar.\label{eq:26}
\end{align}
This is the typical two-qubit i-SWAP gate. With such an universial
gate, assisted by arbitrary rotations of single qubits, any quantum
computing network could be constructed~\cite{Lloyd}.

\section{Discussions and conclusions}
Readout of the qubits is another crucial tasks in quantum computing.
In Barnes et al's scheme~\cite{Barnes}, the flying qubit is encoded
by the spin-states of the electrons in the MQDs and its readout is
implemented by using the usually magnetic Stern-Gerlach effect. In
our proposal the flying qubit is encoded by the lowest two levels of
the electron in the moving trapped potential. These levels are
theoretically steady but still exist weak tunnelings. Thus, by
detecting the tunnelings of the moving electron from the trapped
potential, one can achieve the qubit readouts. This is because that
the tunneling rates of electron in either the state $|0\rangle$ or
the state $|1\rangle$ should be different and thus could be
distinguished individually.
In fact, these tunneling-measurements have been realized in the
recent experiment~\cite{KataokaL}. There, another channel is
introduced to detect the tunnelings of the electrons in the MQDs
across the computational channels. Physically, the flying electron
in the state $|1\rangle$ should yield significantly-high probability
of tunneling to the detecting channel, and thus decrease the current
$I_{\rm top}$ flowing along the computational channel. While, if the
flying electron in the ground level $|0\rangle$, then the
probability of tunneling out should be obviously small and thus
$I_{\rm top}$ should be almost unchanged.
Stronger tunnelings are also possible, if the flying qubit is
excited for leakage. This can be achieved by applying a resonant
pulse to excite the electron staying at the computational basis
$|0\rangle$ (or $|1\rangle$) to the higher level (e.g., the state
$|2\rangle$) with significantly-bigger
tunneling-probabilities~\cite{pepper07}. By this way, flying qubit
staying at $|0\rangle$ or $|1\rangle$ could be more robustly
detected.

Another challenge for realizing our proposal is how to hold only one
electron in a MQD across the computational channel. Initially, many
electrons can be captured by the SAWs from the source region of
2DEG; the number of electrons residing in the minima of SAWs depend
on the size of the formed quantum dot. Note that the static
potential generated by the split-gate is fixed, but the depth and
the curvature of the MQD vary with the time during the MQD moving
along the channel. When the size of the dot becomes smaller,
electrons captured from the source are ejected from the dot and let
a few ones be still trapped by the potential. By suitably
controlling the relevant parameters, e.g., the power of the SAW and
the split-gate voltage, only one electron could reside in a MQD for
realizing the desirable flying qubit. Finally, as in all the other
solid-state quantum computing candidates, decoherence in the present
flying qubit is also an open problem and would be discussed in
future.

In summary, we have put forward an approach to implementing quantum
computation with the energy levels of the electrons trapped in the
MQDs. The idea involves the capture of electrons from a 2DEG by the
SAWs to form the potentials for trapping a single electron. Each SAW
may capture many electrons from the 2DEG source, but we can make
only one electron reside in the minimum of the SAW by tuning the
surface split-gate to change the barrier height, that forces the
excessive electrons to tunnel out from the quantum dot. By numerical
method, we have known that few adiabatic levels of each electron
could be formed in a MQD, and the lowest two ones are utilized to
encode a flying qubit. We have shown how to implement the Rabi
oscillations with the flying qubit for performing single-qubit
operation. A two-qubit gate, i.e., i-SWAP gate, has also be
constructed by using the Coulomb interaction of the electrons in
different MQDs across the nearest-neighbor computational channels.
In principle, our proposal can be extended to the system including
$N$ qubits by integrating an array of $N$ Q1DCs.

\section*{Acknowledgments}

This work was supported in part by the National Science Foundation
grant No. 10874142, 90921010, and the National Fundamental Research
Program of China through Grant No. 2010CB923104, and the Fundamental
Research Funds for the Central Universities No. SWJTU09CX078. We
thank Prof. J. Gao for encouragements and Dr. Adam Thorn for kind
helps.

\end{document}